\documentstyle[manuscript,aps,psfig]{revtex}
\begin{document}

\title{The Faraday quantum clock and non-local photon pair correlations}
\author{Y. Japha$^*$ and G. Kurizki \\ Department of Chemical Physics,
The Weizmann Institute of Science. Rehovot, Israel, 76100}

\maketitle

\begin{abstract}
We study the use of the Faraday effect as a quantum clock for measuring
traversal times of evanescent photons through magneto-refractive structures. 
The Faraday effect acts both as a phase-shifter and as a filter
for circular polarizations. Only measurements based on the Faraday
phase-shift properties are relevant to the traversal time measurements.
The Faraday polarization filtering may cause the loss of non-local
(Einstein-Podolsky-Rosen) two-photon correlations, but this loss
can be avoided without sacrificing the clock accuracy.
We show that a mechanism of destructive
interference between consecutive paths is responsible for superluminal
traversal times measured by the clock.

\end{abstract}

\section{Introduction}

A quantum clock is an observable capable of measuring the time duration of a physical
process in a quantum system\cite{Peres,Baz}. One of the major applications
discussed in this context is the measurement of the traversal time of
a quantum particle through a specific region in 
space\cite{Baz,Buttiker,Sokolovski,Deutsch,Gasparian}: it can
be performed by applying a field in this region, thereby inducing
the evolution of the particle's internal states at a
rate which is a known function of the field strength. If the field is
weak enough, then its effect on the translational degrees of freedom of the
particle is assumed to be negligible and the traversal time can then be
deduced from the final internal state of the particle outside the region
where the field takes effect.

The first quantum clock to be discussed was based on the
Larmor spin precession of an electron in a magnetic 
field\cite{Buttiker,Sokolovski}.
By analogy, it was later proposed to use the Faraday polarization rotation
of an electromagnetic wave in a birefringent crystal\cite{Deutsch}
or in a magneto-refractive medium\cite{Gasparian} as a clock for
photon traversal times in dielectric structures. 
For both Larmor precession and 
Faraday polarization rotation, the theoretical analysis led to the definition 
of a complex time variable\cite{Buttiker,Sokolovski,Gasparian}.
In the case of the Larmor clock for electrons, its real and imaginary parts
measure the spin rotation perpendicular and parallel to the field,
respectively. In the case of the Faraday clock for photons, its real part
measures the rotation of the main axis of polarization while its imaginary
part measures the ellipticity of the resulting polarization.

The works surveyed above have left certain open questions of fundamental 
importance: (i) How do the measured values of traversal times depend
on the clock precision? (ii) What is the origin of the differences between
traversal times measured by quantum clocks through different measuring
schemes and how can they become superluminal\cite{Chiao,Spielman}?
(iii) Whereas field quantization is {\it not}
required for analyzing the evanescent-wave traversal (tunneling) of 
electromagnetic wavepackets through dielectrics or the corresponding 
traversal times,  are there {\it nonclassical} properties of evanescent
photons, which are affected by the precision of their internal quantum clock?
Our purpose here is to elucidate the above questions for the Faraday rotation
clock. 

In Section~\ref{sec:interf} we use our
previously introduced explanation that evanescent-wave transmission or 
tunneling results from {\it destructive interference of internal traversal 
paths}\cite{we}, to show that {\it the
mere possibility of a time-measurement by an internal clock tends to affect
this interference and thus increase the transmission}.
We demonstrate this fundamental result for 
the transmission of photons through a
layered dielectric medium with Faraday rotation. In Section~\ref{sec:ttime} 
we discuss different
measuring schemes and their relevance to the traversal time problem.
We show how superluminal traversal times measured by the clock are caused by
strong destructive interference of traversal paths 
in the evanescent-wave (tunneling) regime.
In Section~\ref{sec:EPR} we discuss the effect of the Faraday 
clock in a dielectric structure on {\it two-photon correlations} and show that 
{\it correlations are not lost by the introduction of a Faraday clock}, unless the
filtering effect of the clock on circular polarizations is significant.  

\section{Path Interference and the Faraday Clock Precision}
\label{sec:interf}

The Faraday effect of polarization rotation in a magneto-refractive medium 
is caused by
the fact that the left-hand ($|+\rangle $) and right-hand ($|-\rangle $) circular
polarization components with respect to an applied magnetic field have
different refractive indices $n_+$ and $n_-$ in the medium. The linear
polarizations $|\leftrightarrow\rangle ,|\updownarrow\rangle $ are superpositions 
of the two circular polarizations with equal
amplitudes. While propagating a distance $x$ through the medium,
the two circular polarization states
$|\pm\rangle \equiv(|\leftrightarrow\rangle \pm i|\updownarrow\rangle )/\sqrt{2}$ of 
light at frequency $\omega$ acquire different phases $\phi_{\pm}=n_{\pm}\omega x/c$.
If the transmission amplitudes of the two circular polarization states
are equal, then a linearly polarized photon entering the medium will exit
with another linear polarization rotated 
by an angle $\theta=\phi_+-\phi_-$ with respect to the
initial one. However, if the transmission amplitudes are
different for the two circular polarizations, then the transmitted photon is
elliptically polarized. In the extreme case where only one
circular-polarization component is transmitted, 
the medium acts as a perfect filter for circular
polarizations.

The use of the Faraday effect as a quantum clock is based on the fact that
the phase difference between the two circular-polarization states $\phi_+ -
\phi_-$ is proportional to the optical length $x\bar{n}$ in the medium, where
$\bar{n}=(n_+ + n_-)/2$ is the mean refractive index.
If the transmission probabilities of the two circular-polarization states
are equal, then the polarization rotation of an initially linearly polarized
photon is proportional to the time it spent in the medium, namely 
\begin{equation} \theta=\omega\frac{n_+-n_-}{\bar{n}}\frac{x}{\bar{v}} \end{equation}
where $\bar{v}=c/\bar{n}$ is the mean velocity in the medium.

Here we discuss the transmission of a photon through a magneto-refractive
periodic dielectric structure ("a photonic band-gap structure").
The structure is made of $N$ alternating dielectric layers $1$ and $2$ with
dielectric constants $n_1,n_2$ and widths $d_1,d_2$, so that the total 
width of the
structure is $L=N(d_1+d_2)/2$.
In the presence of a magnetic field each layer acquires different refractive
indices for left- and right-circular polarizations. The Faraday effect can
measure the traversal time through the structure if the ratio
$\frac{n_+ -n_-}{\bar{n}}$ is kept constant for the two layers.
The transmitted wave at $x=L$ is a superposition of many partial waves
corresponding to different patterns of internal reflections between the layers.
Each such pattern can be described as a path $j$, traversing $N^j_1$ times
the type 1 layers and $N^j_2$ times the type 2 layers, with a certain
number of internal reflections between the layers, each having an
amplitude $\pm\frac{n_1-n_2}{n_1+n_2}$. We can assume that in a weak
magnetic field this inter-layer reflection amplitude is approximately the same
for the two circular polarizations, and therefore the total amplitude of
transmission along a certain path $j$ is equal for the two polarization
states.

\begin{figure}
\psfig{file=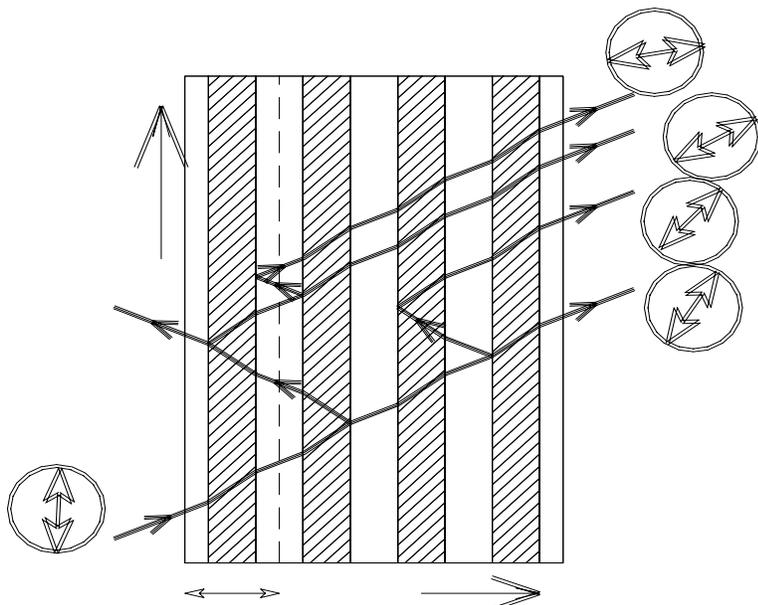,height=8cm,width=10cm,angle=90}
\caption{The Faraday clock for measuring the evanescent-wave traversal time
through a magneto-refractive layered structure. The polarization rotations 
of the different 
interfering paths are proportional to the path length.}
\label{fig:illus}
\end{figure}

It follows from the discussion above, that the
polarization state of an initially linearly polarized photon traversing a
certain path $j$ in the dielectric structure remains linear and is rotated
by an angle
$\theta_j=\Omega(N^j_1\tau_1+N^j_2\tau_2)$, where
$\tau_i=d_i \bar{n}_i/c$ ($i=1,2$) and $\Omega\equiv
2\omega\frac{n_{1+}-n_{1-}}{n_{1+}+n_{1-}}=
2\omega\frac{n_{2+}-n_{2-}}{n_{2+}+n_{2-}}$, 
as illustrated in Fig.~\ref{fig:illus}.
The state of the transmitted photon $|\psi_{tr}(x,t)\rangle $ at $x>L$ is
given by the superposition
\begin{equation} |\psi_{tr}(x,t)\rangle =e^{i\omega(x-L)/c}\sum_j c_j e^{i\omega\tau_j}
|\theta_j\rangle  \label{psitr:exp} \end{equation}
where $c_j$ is the sum over amplitudes for transmission through paths $j$,
which traverse the two different dielectric layers $N^j_1$ and $N^j_2$
times and $\tau_j=N^j_1\tau_1 +N^j_2\tau_2$. The state $|\theta_j\rangle $ 
with $\theta_j=\Omega\tau_j$, is the linear polarization state
\begin{equation} |\theta_j\rangle =\cos\Omega\tau_j|\leftrightarrow\rangle 
+\sin\Omega\tau_j|\updownarrow\rangle 
=\frac{1}{2}\left(e^{-i\Omega\tau_j}|+\rangle +e^{i\Omega\tau_j}|-\rangle \right). 
\end{equation}

A simpler form of the photon states is obtained by using the circular
polarization basis. If the incident photon is in the state
$|\psi_{in}\rangle =|\leftrightarrow\rangle =(|+\rangle +|-\rangle )/\sqrt{2}$,
then the transmitted photon
state is
\begin{equation} |\psi_{tr}(x,t)\rangle =\frac{e^{i\omega(x-L)/c}}{\sqrt{2}}
\left[s(\omega+\Omega)|+\rangle 
+s(\omega-\Omega)|-\rangle \right] \label{psixt} \end{equation}
where $s(\omega)=\sum_j c_j e^{i\omega\tau_j}$ is the spectral transmission
function for a monochromatic wave with frequency $\omega$.

Since the two parts of the wavefunction in Eq.~(\ref{psixt}) are 
distinguishable by a circular polarizer, there is no interference
between them and the total probability of the photon to be transmitted
can be written as an {\it average} of the transmission probabilities
of photons incident on a non-gyrotropic barrier at frequencies $\omega\pm\Omega$
\begin{equation} P_{tr}(\omega)=\frac{1}{2}\left[P_+(\omega) +P_-(\omega)\right] 
\label{Ptrw} 
\end{equation}
where $P_{\pm}=|s(\omega \pm\Omega|^2$. This loss of interference leads to the
enhancement of transmission probability in frequency bands where the
transmission is low. In order to demonstrate this point, notice that
if $\Omega$ is small compared to the scale of variation of $P_0(\omega)$, then
we obtain from Eq.~(\ref{Ptrw})
\[ P_{\Omega}(\omega)\approx P_0(\omega)+\frac{1}{2}\Omega^2
\left. \frac{\partial^2 P_0(\nu)}{\partial \nu^2}\right|_{\nu=\omega} \]
In frequency bands where the transmission is low, especially in the
tunneling (evanescent) regime, the curvature $\partial^2 P_0(\nu)/\partial \nu^2$ is
positive. We therefore find that the presence of the clock raises the transmission
probability (as compared to the same probability in the absence of a clock),
because the destructive interference of traversal paths is 
progressively washed out as $\Omega$ increases.

\section{Time measurements by the Faraday Clock: Superluminality as Path
Interference}
\label{sec:ttime}

What kind of measurement has to be performed in order to obtain
a meaningful traversal time through the dielectric structure in 
Fig.~\ref{fig:illus}?
If we were able to single out only transmitted photons that leave the 
structure in the exact state $|\theta\rangle $, 
then we could know what group of paths those photons went through and
determine the exact traversal time through the dielectric structure. However, 
since the space of polarization states is spanned by only two orthogonal
basis states, we are limited to the analysis of their detection probabilities
rather than the determination of their state. 
In what follows we consider several alternative measuring schemes and discuss
their relevance to the traversal time problem:

A. The first alternative is to measure the number of transmitted
photons $N_{orth}$ with linear polarization orthogonal to the initial
polarization direction and compare it to the total number $N_{tr}$ of
transmitted photons. The traversal time $\tau_{orth}$ can be defined by
assuming that the polarization state of the photon after a time $\tau_{orth}$
is given by $|\theta=\Omega\tau_{orth}\rangle $. This implies
that
\begin{equation} N_{orth}/N_{tr}=\sin^2(\Omega\tau_{orth}). \end{equation}
We then find
\begin{equation} \tau_{orth}=\frac{1}{\Omega}\sin^{-1/2}\left[\frac{|s(\omega+\Omega)
- s(\omega-\Omega)|^2}{4 |s(\omega)|^2}\right] \label{torth} \end{equation}
In the limit $\Omega\rightarrow 0$ this becomes
\begin{equation} \tau_{orth}=\left|\frac{1}{s}\frac{\partial s}{\partial \omega}\right|
=\sqrt{(\partial \log |s|/\partial\omega)^2+(\partial\phi_s/\partial\omega)^2} 
\label{torth1} \end{equation}
where $\phi_s$ is the phase of the transmission function $s=|s|e^{i\phi_s}$.
The variable $\tau_{orth}$ is well known in the context of electron tunneling
through potential barriers as "the B\"uttiker-Landauer 
time"\cite{Buttiker,Review}.

\begin{figure}
\psfig{file=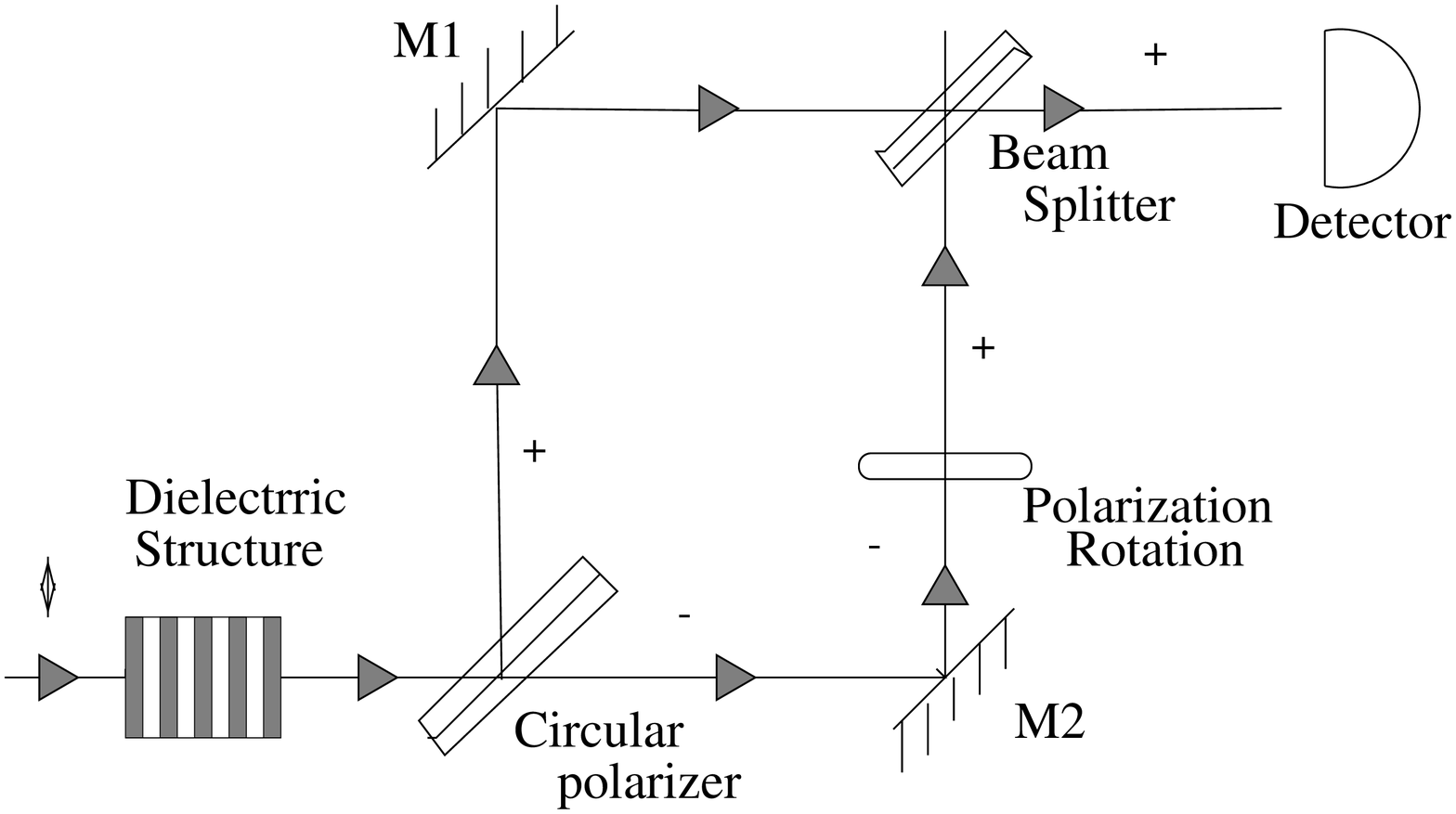,width=12cm,height=8cm}

\caption{Measurement of the phase difference between the two circular 
polarization components $+$ and $-$ of a photon transmitted through a 
Faraday-rotation structure. }

\label{fig:scheme}

\end{figure}

B. The second alternative is to make a direct measurement of the phase 
difference
between the two circular polarization components of the transmitted photons.
This can be done by using a circular polarizer to separate
the two components of the transmitted photons and
then join them again by a beam-splitter as shown in Figure~\ref{fig:scheme}.
If the optical length of one arm of this interferometer with respect to the
other is adjustable, then the detection rate at one port of the beam-splitter
as a function of the length difference between the arms
satisfy the proportionality relation
\begin{equation} P_{det}(\Delta x)\propto \left|s(\omega+\Omega) +is(\omega-\Omega)
 e^{i[\Delta\varphi+\omega\Delta x/c]}\right|^2 \end{equation}
where $\Delta \varphi$ is an additional phase difference introduced by the
optical components. The phase difference between
the two polarization components is then given by
$\phi_+ - \phi_- = \Delta\varphi+\omega \Delta x_{max}/c+\pi/2$, where 
$\Delta x_{max}$
is the value of $\Delta x$ which maximizes $P_{det}(\Delta x)$.
The traversal time in the dielectric structure can then be extracted as
\begin{equation} \tau_{ph}=\frac{\phi_+ -\phi_-}{2\Omega}
\begin{array}{c} \longrightarrow \\ \Omega\rightarrow 0 \end{array}
\frac{\partial \phi_s}{\partial \omega} \label{tph} \end{equation}
which is the well-known value obtained in wavepacket measurements.

\begin{figure}
\psfig{file=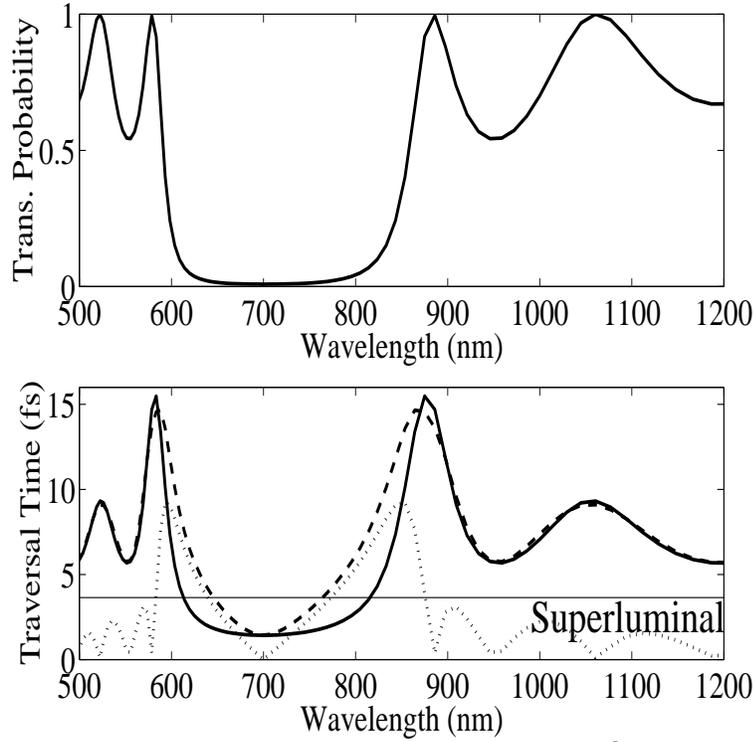,height=10cm,width=10cm}
\caption[]{(a) Transmission probability $|s(\omega)|^2$ for the 
dielectric-layered structure described in 
Ref.~\protect\ref{Chiao}. (b) Traversal times for the same structure: Solid line - 
$\tau_{ph}$ (Eq.~[\protect\ref{tph})]. Dashed - $\tau_{orth}$ 
(Eqs.~[\protect\ref{torth}),(\protect\ref{torth1})]. 
Dotted:  $\tau_{amp}$ (Eq.~[\protect\ref{tamp})].
Note that the center of the band gap is approximately a symmetry point
for all the plotted quantities. }
\label{fig:figuur}
\end{figure}

C. The third alternative is to scan the rotation angles of polarization
from $\theta=0$ to $\theta=\pi$ by a linear polarizer, and define the
peak rotation angle as the polarizer angle where the maximal number of counts
is obtained. The counting
probability for a polarizer at angle $\theta$ is given by
\begin{equation} P(\theta)=\left|\langle \theta|\psi_{tr}\rangle \right|^2=
\left| s_+ e^{i\theta}+s_-e^{-i\theta} \right|^2 \label{ptheta1} \end{equation}
where $s_{\pm}=s(\omega\pm\Omega)$.
The maximal value of this function is obtained at
$\theta=(\phi_+ -\phi_-)/2$, which corresponds to the same phase-time
$t_{ph}$ as given in Eq.~(\ref{tph}). In a periodically-layered
dielectric structure, the evanescent-wave transmission probability
$|s(\omega)|^2$ is nearly symmetric with respect to the center of 
the forbidden band gap (Fig.~\ref{fig:figuur}(a)) and so are the 
traversal times (\ref{torth1}) and (\ref{tph}) (Fig.~\ref{fig:figuur}(b)).
Hence, $|s_{+}| \approx |s_{-}|$ if $\omega$ is at the band-gap center.

\begin{figure}
\psfig{file=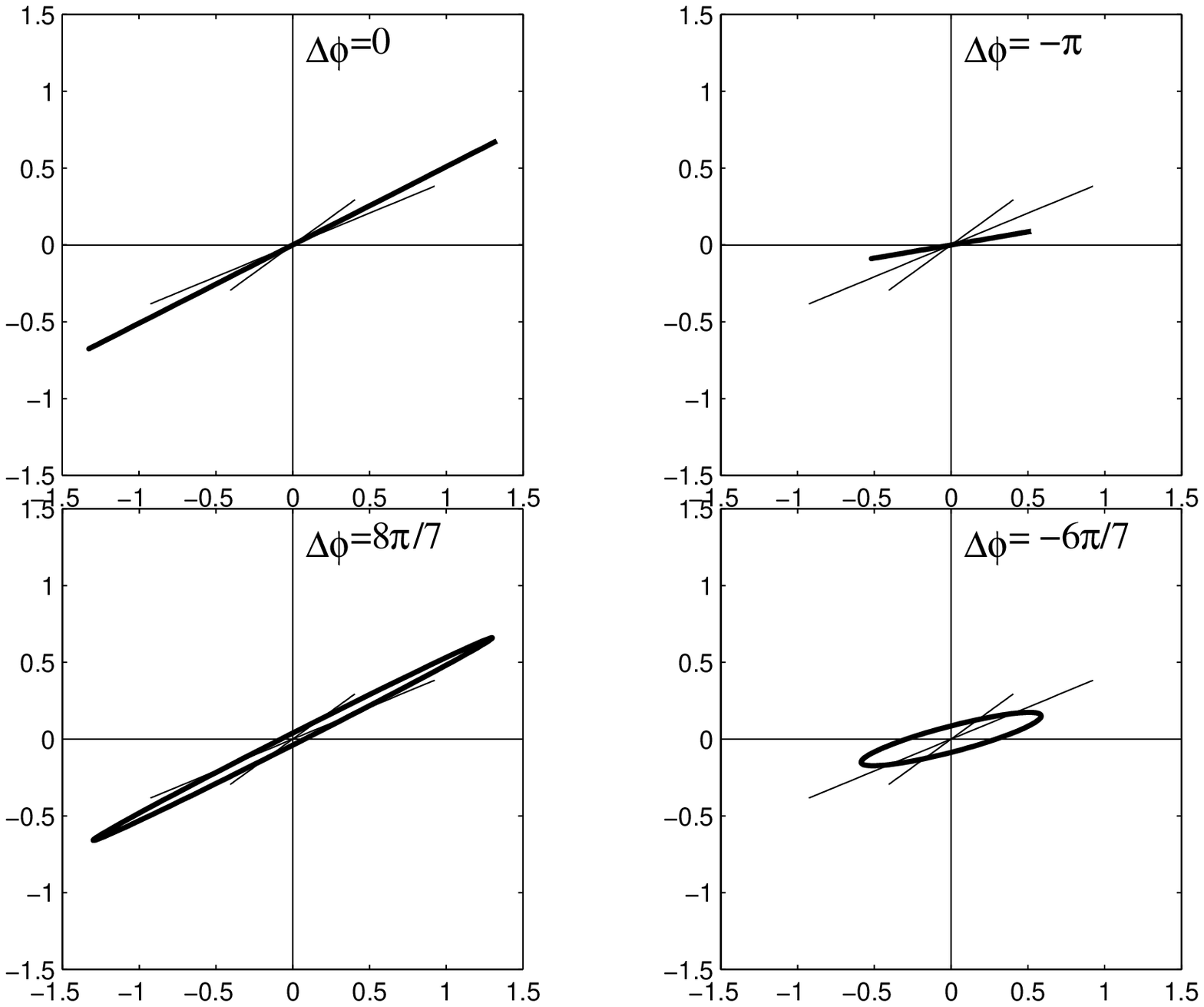,width=11cm,height=8cm}
\caption{A superposition $|\theta_{12}\rangle$ (thick line) of two linear
polarization states $|\theta_1\rangle , |\theta_2\rangle $ (thin lines) with 
different relative phases $\Delta\phi$: (a) $\Delta\phi=0$ 
yields $\theta_{12}>\theta_2>\theta_1$ with large amplitude; 
(b) $\Delta\phi=\pi$ yields $\theta_{12}<\theta_1<\theta_2$ with small 
amplitude; (c), (d) $\Delta\phi$ that is a non-integer fraction of $\pi$ yields 
elliptically polarized states.}
\label{polfig}
\end{figure}

In Ref.~\ref{we} we have shown that low transmission of a wavepacket 
together with abnormally
short wavepacket peak traversal time is a consequence of {\it destructive 
interference} between consecutive partial transmitted wavepackets. In order
to demonstrate the effects of interference on the clock read-out,
we analyze the polarization state formed by a superposition of
two linear polarization states that appear in Eq.~(\ref{psixt}). Consider the
polarization state 
\begin{equation} |ij\rangle \equiv \lambda_i |\theta_i\rangle
+\lambda_j |\theta_j\rangle , \end{equation} 
where $\lambda_j\equiv c_j e^{i\omega\tau_j}$ is the
amplitude of the $j$th transmitted wave. As demonstrated in 
Figure~\ref{polfig}, if
$\theta_i<\theta_j$ and $|\lambda_i|>|\lambda_j|$, it can be shown that if the
interference between the terms $i$ and $j$ is predominantly destructive,
i.e., when $\Re\{\lambda^*_i\lambda_j\}<0$, then the main axis of polarization of
the resulting elliptic polarization state $|ij\rangle $ has an angle
$\theta_{ij}<\theta_i$. In
particular, if the phase difference between $\lambda_i$ and $\lambda_j$ is exactly
$\pi$, such that $\lambda_j/\lambda_i=-r$, then the resulting state
has linear polarization
\[ \theta_{ij}=\theta_i
-\arctan\left[\frac{r\sin(\theta_j-\theta_i)}{1-r\cos(\theta_j-\theta_i)}
\right] \]
and $\theta_{ij}<\theta_i<\theta_j$ whenever $r\cos(\theta_j-\theta_i)<1$.

In general, the main axis of the elliptic polarization state of the
transmitted photons is given
by the maximum overlap with a linear polarization state $|\theta_m\rangle $.
Using Eq.~(\ref{psitr:exp}), we can express the probability function in
Eq.~(\ref{ptheta1}) as
\begin{equation} P(\theta)=\left|\sum_j c_j e^{i\omega\tau_j} 
\cos(\theta-\theta_j)\right|^2
\end{equation}
In order to obtain the value of $\theta_m$ we find the maximum of $P(\theta)$
by equating its derivative to zero.  We then obtain
the following angle of the main axis of polarization
\[ \tan 2\theta_m=\frac{\sum_j c_j^2\sin 2\theta_j
+\sum_{i\neq j}c_i c_j e^{-i\omega(\tau_i-\tau_j)}\sin(\theta_i+\theta_j)}
{\sum_j c_j^2\cos 2\theta_j+\sum_{i\neq j} c_i c_j e^{-i\omega(\tau_i-\tau_j)}
\cos(\theta_i+\theta_j)} \]
If $\Omega$ is very small, then  $\sin 2\theta\approx 2\theta$ and
$\cos 2\theta\approx 1$. The clock time then reads exactly the same as the
mean wavepacket traversal time\cite{we}. Hence, {\it destructive interference} 
between different terms $i\neq j$ gives rise to predominantly negative
contributions $c_i c_j e^{-i\omega(\tau_i-\tau_j)}$, which {\it cause the
effect of abnormally short or even superluminal clock read-out times}.

\section{Faraday Clock Effect in Nonlocal (EPR) Photon Correlations}
\label{sec:EPR}

\begin{figure}
\vspace{0.5cm}
\psfig{file=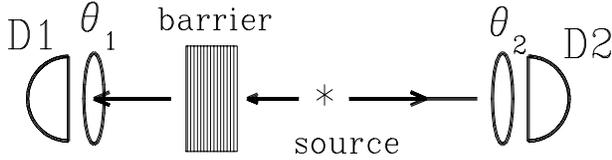,angle=90,width=8cm,height=2cm}
\vspace{0.5cm}
\caption{Measurement of two-photon EPR correlations when
one photon is transmitted through a Faraday-rotating reflective structure.}
\label{fig6}

\end{figure}

Here we consider measurements that have never been discussed before, which
pertain to non-local (Einstein-Podolsky-Rosen) correlations between entangled photons.
Suppose that we repeat the Aspect experiment\cite{Aspect} using a correlated 
pair of photons, but insert a Faraday-rotating reflective structure (of the 
kind depicted in Fig.~(\ref{fig:illus})) in the way of 
one of the photons (Figure~\ref{fig6}).
If the correlations between the transmitted photon and its twin are
reduced, then the violation of Bell's inequality is expected to be 
weakened. To what extent does the Faraday clock affect the violation of the
inequality?

Let us take the initial state of the field to be
\begin{eqnarray}
 |\phi_{in}\rangle &=&\frac{1}{\sqrt{2}}(|\updownarrow,\updownarrow\rangle 
+|\leftrightarrow,\leftrightarrow\rangle )= \nonumber
 \\
&&=\frac{1}{\sqrt{2}}(|+,-\rangle +|-,+\rangle )
\end{eqnarray}
where the first entry in the ket-vector refers to the state of the
photon transmitted through the Faraday-rotating structure and travelling 
to detector 1 and the second entry refers to the
state of the photon travelling to detector 2. 

Bell's inequality in the formulation of Clauser-Horne-Shimony-Holt
(CHSH) makes use of the correlation function\cite{EPR}
\begin{equation} E(\theta_1,\theta_2)=
\frac{\langle (I^1_{\updownarrow}-I^1_{\leftrightarrow})
(I^2_{\updownarrow}-I^2_{\leftrightarrow})\rangle }
{\langle (I^1_{\updownarrow}+I^1_{\leftrightarrow})
(I^2_{\updownarrow}+I^2_{\leftrightarrow})\rangle} 
\label{CHSH} \end{equation}
where $I$ is the measured intensity, the superscripts refer to intensity 
measured at detectors 1 and 2 and the signs $\updownarrow,\leftrightarrow$ refer to 
measurements with polarizer set at angles $\theta_i$ and $\theta_i+\pi/2$
($i=1,2$).

Since detector 1 measures only the fraction of photons transmitted through
the Faraday-rotating structure, the state of the photons near the detectors is
given by $|\phi_{trans}\rangle =\frac{1}{\sqrt{2}}
(s_+|+,-\rangle +s_-|-,+\rangle )$. We then
find
\begin{eqnarray}
 \langle I^1_{\updownarrow} I^2_{\updownarrow}\rangle
&=&|\langle \phi|\theta_1,\theta_2\rangle |^2=
 \nonumber \\
 &&=\frac{1}{2}| s_+ e^{-i\psi}+s_- e^{i\psi}|^2 \nonumber \\
 \langle I^1_{\updownarrow} I^2_{\leftrightarrow}\rangle &=&
\frac{1}{2}|s_+ e^{-i\psi}-s_- e^{i\psi}|^2
\end{eqnarray}
where $\psi=\theta_1-\theta_2$, and similarly for the other correlations.
The correlation function in Eq.~(\ref{CHSH}) becomes
\begin{equation} E(\theta_1,\theta_2)=\frac{2|s_+||s_-|}{|s_+|^2+|s_-|^2}
\cos(2\psi+\phi_+-\phi_-) \label{E12} \end{equation}

The CHSH inequality for measurements at polarizer angles  
$\theta_i,\theta'_i$ is
\[ |B| =|E(\theta_1,\theta_2)-E(\theta_1,\theta'_2)+E(\theta'_1,\theta'_2)- 
E(\theta'_1,\theta_2)| \leq 2 \]

\begin{figure}
\psfig{file=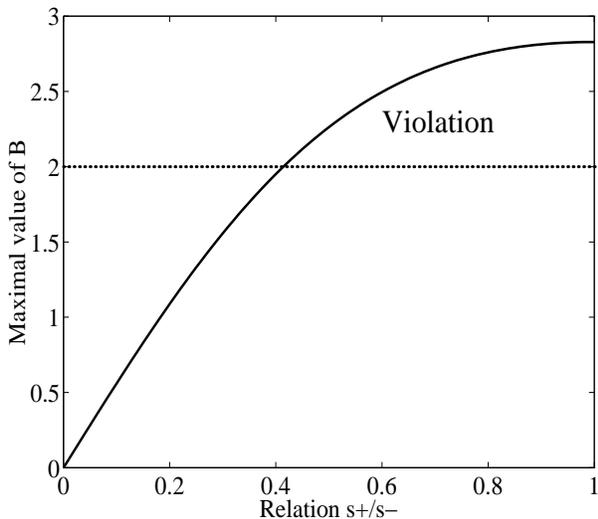,width=8cm,height=7cm}
\caption{The maximal value of the factor B for various values of
the clock precision $\Omega$ when one of the correlated photons
is transmitted through a dielectric layered structure as in Fig.~\ref{fig:figuur}.}
\label{demons}
\end{figure}

Without a Faraday rotating reflective structure we would have had 
$E(\theta_1,\theta_2)=\cos(2\psi)$ but 
with a barrier the cosine term is
shifted by a certain phase and multiplied by a factor which is {\it 
always smaller or equal to unity} (Fig.~\ref{demons}). 
This factor is maximal when $|s_-|=|s_+|$, namely, when the right- and the 
left-circular polarizations are not separated. This happens if $\Omega$ is 
either very small (weak clock) or when $\omega$ {\it lies at a symmetry point} 
(extremum) of  the $|s(\omega)|^2$ curve, e.g.,  
if $\omega$ lies in {\it the center of the band gap} of the structure 
(Fig.~\ref{fig:figuur}(a)).
By contrast, if the two polarizations are not equivalent, 
$E(\theta_1,\theta_2)$ 
is reduced and {\it the violation of the inequality is diminished}, indicating 
partial loss of the two-photon correlations. This loss of correlation is 
explained by the fact that the Faraday-rotating structure acts as a filter 
for certain polarizations 
and thus the transmitted photons measured at detector 1 are a sub-ensemble 
of the original photons, which does not represent the polarization state 
of the initial photon population.
When $\Omega$ is small, it is easy to show from Eq.~(\ref{E12}) that 
the maximal value of $E(\theta_1,\theta_2)$ becomes
\begin{equation} E_{max}\approx 1-2\Omega^2\tau_{amp}^2 \end{equation}
where 
\begin{equation} \tau_{amp}=\frac{1}{|s|}\frac{\partial |s|}{\partial \omega}
=\frac{\partial}{\partial\omega}\log |s| 
\label{tamp} \end{equation}
The variable $\tau_{amp}$ measures the effectiveness of the Faraday-rotating
structure as a circular polarizer. It measures the ability of the 
structure to separate between left- and right-circularly 
polarized photons. 
Although having units of time, it is difficult to refer to $\tau_{amp}$
any meaning connected with time duration. However, it is possible to treat 
$\tau_{amp}^{-1}$ as a typical frequency scale over which the transmissivity 
of the dielectric barrier is significantly changed. 

\section{Summary}

We have analysed in this paper the transmission of photons through a 
layered dielectrtic medium with Faraday rotation and have sketched measurements 
of their traversal times in order to elucidate possible quantum clock 
mechanisms in the evanescent-wave (tunneling) regime. For the first time, 
both single-photon and two-photon (quantum electrodynamical) properties
have been considered in this context. 

Destructive interference between different traversal paths has been shown to
be the origin of the effect of abnormally short or even superluminal clock 
read-out times. Several possible schemes have been proposed for traversal-time
measurements based on the Faraday-clock phase shifts.

In frequency bands where the transmission is low, especially in the
tunneling (evanescent-wave) regime, we have shown that the presence of the 
clock tends to wash out the destructive interference and thereby raise the 
transmission probability as compared to its counterpart in the absence of 
a clock.

Perhaps the most intriguing subject considered here are measurements  of 
non-local correlations between two entangled photons with a Faraday-rotation 
structure in the way of one of the photons. 
The two-photon correlation function $E(\theta_1,\theta_2)$, which is used to
calculate the violation of Bell's inequality, is shifted by a phase and
multiplied by a factor which is always smaller or equal to unity 
(Figure~\ref{demons}), as a result of the Faraday rotation. This factor is
maximal when the right and left circular polarizations are not separated. 
This happens if the Faraday-clock rate is either very small or if the 
photon frequency $\omega$ lies at a symmetry point (extremum) of the $s(\omega)$ 
transmission curve, e.g., if $\omega$ lies at the center of a band gap.
The latter result is remarkable, since it allows us to measure two-photon
entanglement (Bell's inequality violation) without sacrificing the Faraday
clock accuracy.    
By contrast, if the two polarizations are not equivalent, 
$E(\theta_1,\theta_2)$ 
is reduced and {\it the violation of the inequality is diminished}, indicating 
partial loss of the two-photon correlations. This loss of correlation is 
explained by the fact that the Faraday-rotating structure acts as a filter 
for certain polarizations 
and thus the transmitted photons measured at detector 1 are a sub-ensemble 
of the original photons, which does not represent the polarization state 
of the initial photon population.

{\it $^*$ Present address: Clarendon Laboratory, Department of Physics,
 University of Oxford, Oxford, UK OX1 3PU}
 
\section{Acknowledgments}
This work has been supported by Minerva, ISF and EU (TMR) grants.

\end{document}